\documentclass[aps,prl,twocolumn,superscriptaddress,longbibliography,floatfix]{revtex4-2}

\usepackage{graphicx}
\usepackage{amsmath,amssymb}
\usepackage{bm}
\usepackage{hyperref}
\usepackage{placeins}
\usepackage{xcolor}

\graphicspath{{figures/}}

\begin{document}

\title{Magnetic Order from First-Principles Linear Response}

\author{Yaroslav Zhumagulov}
\email{yaroslav.zhumagulov@epfl.ch}
\affiliation{Institute of Physics, \'Ecole Polytechnique F\'ed\'erale de Lausanne (EPFL), CH-1015 Lausanne, Switzerland}

\author{Emiliano Cruz-Aranda}
\affiliation{Institute of Physics, \'Ecole Polytechnique F\'ed\'erale de Lausanne (EPFL), CH-1015 Lausanne, Switzerland}

\author{Denis Kochan}
\affiliation{Department of Physics and Center for Quantum Frontiers of Research and Technology (QFort), National Cheng Kung University, Tainan 70101, Taiwan}
\affiliation{Institute of Physics, Slovak Academy of Sciences, 84511 Bratislava, Slovakia}

\author{Oleg V. Yazyev}
\affiliation{Institute of Physics, \'Ecole Polytechnique F\'ed\'erale de Lausanne (EPFL), CH-1015 Lausanne, Switzerland}

\date{\today}

\begin{abstract}
The magnetic order of a material is usually predicted by comparing the energies of trial spin configurations, a strategy that cannot reach orders outside the chosen set and rapidly becomes intractable as the number of magnetic sites grows.
We show that the order can instead be identified directly from the paramagnetic state: the leading eigenmodes of its self-consistent magnetization response, obtained with spin--orbit coupling at arbitrary wavevector within the primitive cell, determine both the ordering wavevector and the pattern of moments, with no candidate configurations.
With no prior assumption about the order, the method correctly reproduces the G-type stacking with out-of-plane moments in the altermagnet candidate KV$_2$Se$_2$O, the $3{:}1$ ferrimagnetic tetrahedra of Cu$_2$OSeO$_3$, and the incommensurate spiral instability of monolayer NiI$_2$.
Our results establish the response of the paramagnetic state as a predictive and unbiased route to assessing magnetic order, enabling systematic searches for complex magnets.
\end{abstract}

\maketitle

Predicting magnetic order remains a challenging task in computational condensed-matter physics~\cite{Olsen2024AFM,PasquierYazyev2022}.
The relative arrangement of magnetic moments and their magnitude determine a magnet's properties: for instance, skyrmions emerge in chiral magnets~\cite{Muehlbauer2009,NagaosaTokura2013}, modulated magnetic structures can transform an insulator into a ferroelectric~\cite{Kimura2003,CheongMostovoy2007}, and van der Waals magnets may retain or alter their magnetic order when the sample thickness is reduced to a few atomic layers~\cite{HuangCrI3_2017,GongCr2Ge2Te6_2017}.
Although symmetry analysis classifies the possible magnetic orders of a given crystal~\cite{PerezMatoMagCryst2015}, it does not determine which specific type is realized in practice.
Consequently, catalogs of magnetic topological materials based on first-principles calculations~\cite{XuMagneticTopological2020} still rely on experimentally refined structures~\cite{GallegoMAGNDATA2016}, thereby limiting their scope to compounds for which the magnetic order has already been determined empirically.

The standard approach in first-principles calculations compares the total energies of predefined ferromagnetic and antiferromagnetic configurations.
While the approach is suitable for high-throughput screening~\cite{Horton2019,TorelliOlsen2020,FreyHighThroughput2020}, it evaluates only the variants selected by the researcher.
Moreover, the number of collinear configurations alone grows exponentially with the number of magnetic sites~\cite{AsadaBlugel1997}.
Symmetry-aware exhaustive search methods can cover all 72,075 nonmagnetic stoichiometric structures from the Inorganic Crystal Structure Database~\cite{Zagorac2019}, but only if the initial analysis is restricted to collinear configurations~\cite{HuangCollinear2026}.
A study of transition-metal dihalide monolayers comparing ferromagnetic and stripe orders found the ferromagnetic state to be energetically favorable for NiI$_2$~\cite{BotanaNorman2019}, whereas the monolayer actually exhibits spiral order~\cite{SongNiI2_2022,MiaoNiI2_2025}.
In a benchmark study of 122 materials using symmetry-selected noncollinear magnetic configurations, the experimentally determined magnetic space group was identified among the calculated states for 90\% of the materials.
However, it corresponded to the minimum-energy state in only 38\%~\cite {Huebsch2021}.

Genetic algorithms~\cite{ZhengMagGene2021}, iterative searches for spin-wave stability~\cite{TellezMoraGroundState2024}, machine-learning-based classifiers~\cite{FreyHighThroughput2020,MerkerMLMagnetism2022}, and mappings to spin Hamiltonians~\cite{LKAG1987,Szilva2023} extend the scope of research.
However, they all start from predefined ordered states rather than the symmetric state from which order emerges.
The issue is particularly acute for unit cells with many magnetic sites~\cite{HobbsAlphaMn2003}, in cases of nearly degenerate ordering patterns, and in situations where symmetry leaves the ordering wavevector undefined.
Comparing energies allows one to determine which investigated state has the lowest energy, but it fails to reveal instabilities outside the chosen set of states or the proximity of a competing ordering channel.

Spin-density functional theory in the noncollinear approximation allows each magnetic moment to have its own direction~\cite{KublerHockSticht1988}, while the generalized Bloch theorem permits the existence of spin spirals with an arbitrary wavevector within the unit cell~\cite{Sandratskii1991,Sandratskii1998,Halilov1998,Rosengaard1997,KurzBlugel2004,HeideBlugel2009,Jacobsson2013}.
Consequently, spiral energy minimization in reciprocal space has been performed for extensive sets of two-dimensional materials~\cite{SodequistOlsen2023,C2DB2024}.
However, spin--orbit coupling (SOC) couples spins to the crystal lattice and invalidates the theorem~\cite{Sandratskii1998}.
Therefore, it is typically treated as an additional perturbation~\cite{SodequistOlsen2023,HeideBlugel2008}.
A single spiral cannot describe an instability characterized by varying amplitudes and directions at different lattice sites, as in B20 FeGe, where the spirals on its four Fe sublattices differ in phase and rotation axis~\cite{GrytsiukBlugel2021}.

Another approach infers the order from the properties of the paramagnetic state.
The Stoner criterion~\cite{Stoner1938} links the onset of ferromagnetic instability to the divergence of the spin susceptibility.
Its generalization to the case of a finite wavevector describes spin-density waves~\cite{Overhauser1962} (e.g., in chromium), where the ordering wavevector is determined by the wavevector dependence of the susceptibility~\cite{Staunton1999}.
When SOC is taken into account, the paramagnetic susceptibility becomes anisotropic, leading to the emergence of an easy axis of magnetization~\cite{MatsumotoStauntonStrange1990}.
The theory of disordered local moments identifies ordering tendencies, including long-period states, from the paramagnetic state of metals with localized moments~\cite{Gyorffy1985,MendiveTapiaStaunton2017,MendiveTapiaStaunton2019}. 

Here we use the response function~\cite{Savrasov1998,SkovhusOlsen2021,Cao2018,GorniTimrovBaroni2018,GorniTurboMagnon2022} to compute the order directly.
Drawing on susceptibility analysis results from low-energy models~\cite{ZhumagulovPRL2024,ZhumagulovPRB2024,HerasymchukPRB2026}, we calculate the leading eigenmodes of the product of the static Kohn--Sham response function $\chi^0(\mathbf q,\omega=0)$, at wavevector $\mathbf q$ and zero frequency, and the Hartree--exchange--correlation kernel $K$.
The product linearizes the self-consistency equations for the paramagnetic state, which is calculated from first principles including SOC.
An eigenvalue exceeding unity indicates the presence of an instability, while the corresponding eigenvector defines the configuration of magnetic moments within the unit cell.
We test the method on three materials, each presenting specific challenges for conventional methods: for the altermagnet candidate KV$_2$Se$_2$O~\cite{JiangKV2Se2O2025,SunKV2Se2O2025}, the key question concerns the stacking of identical Néel-ordered planes, which determines whether the bulk crystal is altermagnetic or not.
In the case of Cu$_2$OSeO$_3$~\cite{BosCu2OSeO3_2008,AdamsCu2OSeO3_2012}, we examine the moment configuration within a large 56-atom unit cell.
Finally, for the NiI$_2$ monolayer~\cite{Kuindersma1981,KurumajiNiI2_2013,SongNiI2_2022,AminiNiI2_2024,MiaoNiI2_2025}, the question is the wavevector of the helical magnetic structure.

\textit{Method}---The central object of our method is the following linearized eigenvalue problem at wavevector $\mathbf q$, from which we determine the largest eigenvalue $\lambda_{\mathbf q}$ and its corresponding eigenvector $\phi_{\mathbf q}$:
\begin{equation}
  \lambda^{\nu}_{\mathbf q}\,\phi^{\nu}_{\mathbf q\alpha}(\mathbf r)=\sum_{\beta\gamma}
  \chi^0_{\alpha\beta}(\mathbf q,\omega=0)\,
  \big[K^{\beta\gamma}\phi^{\nu}_{\mathbf q\gamma}\big](\mathbf r),
  \label{eq:detector}
\end{equation}
using a symmetry-preserving paramagnetic Kohn--Sham reference including SOC, with modes $\nu$ ordered by decreasing $\lambda^{\nu}_{\mathbf q}$.
The components $\alpha=0,x,y,z$ denote the charge density and three magnetization densities.
The fluctuation $\phi_{\mathbf q}$ is density-like, and its magnetic components give the spatial pattern of the incipient order.
The fluctuation combines its two Fourier components as
$\delta n_{\mathbf q\alpha}(\mathbf r)=\phi_{\mathbf q\alpha}(\mathbf r)e^{i\mathbf q\cdot\mathbf r}+\phi^*_{\mathbf q\alpha}(\mathbf r)e^{-i\mathbf q\cdot\mathbf r}$,
where the wavevector $\mathbf q$ fixes the phase relation between unit cells, while the amplitudes, phases, and directions within $\phi_{\mathbf q}$ determine the pattern inside each unit cell.
We can therefore represent long-wavelength or incommensurate order without enlarging the reference cell.

The condition $\lambda_{\mathbf q}^{\nu}=1$ is the instability threshold, and $\lambda_{\mathbf q}^{\nu}>1$ indicates that the spin-symmetric reference is unstable toward the magnetic fluctuation.
The dominant eigenvalue at each $\mathbf q$ identifies the strongest instability, while subleading eigenmodes expose competing orders at the same wavevector.
Comparing the dominant eigenvalues at different $\mathbf q$ then identifies the leading modulation of the symmetric state. 
Collinear order, amplitude-modulated order, and noncollinear or spiral textures therefore appear as different eigenmodes of the same operator rather than as separate ansatzes.

The kernel converts $\phi_{\mathbf q}$ into the perturbing field
$V_{\mathbf q\alpha}=\sum_\beta K^{\alpha\beta}\phi_{\mathbf q\beta}$, where
$K^{\alpha\beta}=\delta^2E_{\mathrm{Hxc}}/\delta n_\alpha\delta n_\beta$ is the second derivative of the Hartree and exchange--correlation energy.
The corresponding matrix-valued perturbation is $\mathcal V_{\mathbf q}=\sum_\alpha\sigma_\alpha V_{\mathbf q\alpha}$, where $\sigma_0$ is the identity and $\sigma_{x,y,z}$ are Pauli matrices.
For local exchange--correlation functionals, rotational invariance at zero magnetization gives
\begin{equation}
  K\big|_{\mathbf m=0}=\mathrm{diag}\big(v+f^{\mathrm c}_{\mathrm{xc}},
  f^{\mathrm s}_{\mathrm{xc}},f^{\mathrm s}_{\mathrm{xc}},f^{\mathrm s}_{\mathrm{xc}}\big),
  \label{eq:kernel_m0}
\end{equation}
where $v$ is the Coulomb interaction, while $f^{\mathrm c}_{\mathrm{xc}}$ and $f^{\mathrm s}_{\mathrm{xc}}$ are the charge and magnetic second derivatives of the exchange--correlation energy, respectively.
The mixed charge--magnetization derivatives therefore vanish, while rotational invariance makes the three magnetic entries equal.
The Coulomb term acts only on the charge channel, while the magnetic block provides the exchange enhancement of the spin response~\cite{Savrasov1998,KatsnelsonLichtenstein2004}.
Although the four-component formulation allows a magnetic mode to carry any charge redistribution returned by $\chi^0$, at the time-reversal-symmetric reference $\chi^0$ returns none, decoupling charge and magnetization.
The roles of the two factors in Eq.~(\ref{eq:detector}) are therefore distinct: $K$ generates the self-consistent field associated with a fluctuation, and $\chi^0$ propagates that field through the electronic structure.
As a result, their product determines whether the fluctuation is enhanced or suppressed.
SOC enters the reference Hamiltonian and the response, while the kernel in Eq.~(\ref{eq:kernel_m0}) remains isotropic in spin space.

\begin{figure*}[!t]
\includegraphics[width=\textwidth]{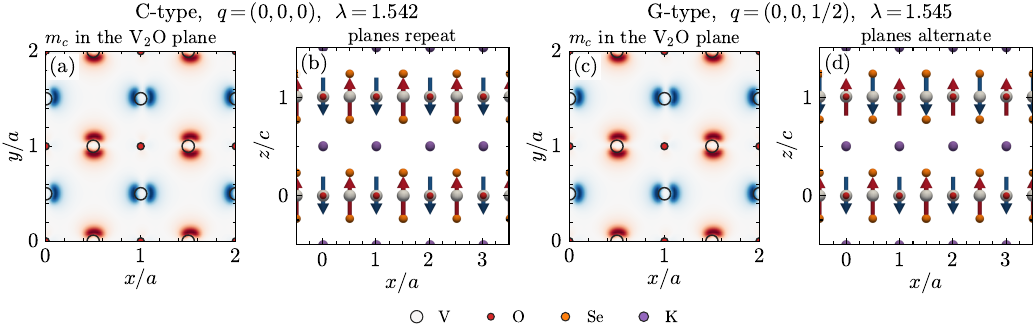}
\caption{\label{fig:kv}%
Leading KV$_2$Se$_2$O response eigenvectors for C-type stacking at $\mathbf q=0$ in panels a and b and G-type stacking at $\mathbf q=(0,0,\tfrac12)$ in panels c and d.
Panels a and c show the $c$-axis magnetization $m_c$ in one V$_2$O plane, with red and blue denoting opposite signs.
Panels b and d show the corresponding V moments across planes, which repeat for C-type stacking and alternate for G-type stacking.
Here $a$ and $c$ are the lattice constants.
The eigenvalue above each pair of panels is $\lambda_c$ of Table~\ref{tab:kv-lambdas}.}
\end{figure*}

The second step is the response to $\mathcal V_{\mathbf q}$.
We evaluate the action of $\chi^0$ with the standard metallic Sternheimer construction~\cite{deGironcoli1995,BaroniRMP}.
For every occupied or partially occupied state, its resonant component satisfies
\begin{equation}
  \big(H_{\mathbf k+\mathbf q}+Q_{\mathbf k+\mathbf q}-\varepsilon_{n\mathbf k}\big)
  |\Delta u^{+\mathbf q}_{n\mathbf k}\rangle
  =-\big(\tilde\theta_{F,n}-P_{\mathbf k+\mathbf q}\big)
  \mathcal V_{\mathbf q}|u_{n\mathbf k}\rangle,
  \label{eq:sternheimer}
\end{equation}
with the Hermitian counterpart at $\mathbf k-\mathbf q$ providing the antiresonant component.
Here $\Delta u^{\pm\mathbf q}_{n\mathbf k}$ are the first-order Kohn--Sham responses.
The Hamiltonian $H_{\mathbf k}$, state $u_{n\mathbf k}$, and eigenvalue $\varepsilon_{n\mathbf k}$ are those of the paramagnetic reference.
The labels $n$ and $\mathbf k$ denote the band and Bloch wavevector, and $Q$, $P$, and $\tilde\theta_F$ are the standard metallic regularization, projector, and occupation factor, respectively~\cite {deGironcoli1995,BaroniRMP}.
The two equations represent the $\mathbf q$ and $-\mathbf q$ components required by a Hermitian perturbation and are linear in the kernel field generated from $\phi_{\mathbf q}$.
Finally, we produce the induced four-component density
\begin{equation}
  \bar n_{\mathbf q\alpha}(\mathbf r)=\frac{1}{N_k}\sum_{n\mathbf k}
  \left[
  u^\dagger_{n\mathbf k}\sigma_\alpha\Delta u^{+\mathbf q}_{n\mathbf k}
  +\Delta u^{-\mathbf q\dagger}_{n\mathbf k}
  \sigma_\alpha u_{n\mathbf k}
  \right],
  \label{eq:induced_density}
\end{equation}
where $N_k$ is the number of sampled $\mathbf k$ points and $\sigma_\alpha$ selects the charge or magnetic component.
Each term combines an unperturbed state with its first-order response.
Summing them reconstructs the induced charge for $\alpha=0$ and the three induced magnetization components for $\alpha=x,y,z$.
The induced density is $[\chi^0(\mathbf q,0)V_{\mathbf q}]_\alpha$ and therefore closes Eq.~(\ref{eq:detector}).
Ultimately, the sequence $\phi_{\mathbf q}\xrightarrow{K}V_{\mathbf q}\xrightarrow{\chi^0}\bar n_{\mathbf q}$ represents the repeated action required to obtain the dominant eigenpairs without constructing $\chi^0$ as a two-point function.

The proximity of eigenvalues indicates the presence of competing tendencies, while the magnitudes of the moments and the energies of the ordered states require self-consistent calculations, which the eigenvectors can seed.

\textit{KV$_2$Se$_2$O: antiferromagnetic order}---The layered oxyselenide KV$_2$Se$_2$O has been characterized as a metallic altermagnet with $d$-wave symmetry at room temperature~\cite{JiangKV2Se2O2025}, and two magnetic stackings, C-type and G-type, have been proposed.
Both stackings are characterized by a checkerboard Néel order within each plane with magnetic moments aligned along the $c$-axis, differing only in the relative arrangement of the layers: in C-type stacking, a 90\textdegree rotation, without translation or inversion, swaps the spin sublattices and allows the $d$-wave spin splitting~\cite{Smejkal2022b,JiangKV2Se2O2025}, whereas in G-type stacking, the sublattices are connected by a lattice translation, which precludes the existence of bulk altermagnetism~\cite{SunKV2Se2O2025}.
Neutron diffraction data confirm the presence of G-type order and indicate that both stacking types are virtually degenerate in energy~\cite{SunKV2Se2O2025}.

We treat the two stackings as two wavevectors: $\mathbf q=0$ for C-type and $\mathbf q=(0,0,\tfrac12)$ for G-type stacking.
At both wavevectors, our method yields the in-plane Néel checkerboard polarized along $c$ (Fig.~\ref{fig:kv}).
The eigenvalues $\lambda^{\mathrm C}_c=1.5417$ and $\lambda^{\mathrm G}_c=1.5452$ (Table~\ref{tab:kv-lambdas}) indicate that G-type stacking is more favorable.
Self-consistent calculations of the two ordered states also favor the same stacking, with a difference of $2.43$~meV per formula unit~\cite{suppl}, consistent with the results of earlier calculations~\cite{Thapa2026}.

\begin{table}[b]
\caption{\label{tab:kv-lambdas}%
KV$_2$Se$_2$O response eigenvalues for C-type stacking at $\mathbf q=0$ and G-type stacking at $\mathbf q=(0,0,\tfrac12)$: the $c$-polarized eigenvalue $\lambda_c$, either $\lambda_{ab}$ of the degenerate in-plane pair, and the SOC splitting $\Delta\lambda_c=\lambda_c-\lambda_{ab}$.}
\begin{ruledtabular}
\begin{tabular}{lccc}
stacking & $\lambda_c$ & $\lambda_{ab}$ & $\Delta\lambda_c$ \\
\hline
C-type & 1.541742 & 1.541243 & $4.99\times10^{-4}$ \\
G-type & 1.545250 & 1.544730 & $5.20\times10^{-4}$ \\
\end{tabular}
\end{ruledtabular}
\end{table}

SOC splits each of the leading triplets into a mode polarized along the $c$-axis and a pair of in-plane states.
The splitting values $\Delta\lambda_c$ listed in Table~\ref{tab:kv-lambdas} favor the orientation of magnetic moments along the $c$-axis.
Thus, the response reproduces the G-type structure with moments along the $c$-axis refined by neutron diffraction~\cite{SunKV2Se2O2025}.

\begin{figure*}[!tbp]
\includegraphics[width=\textwidth]{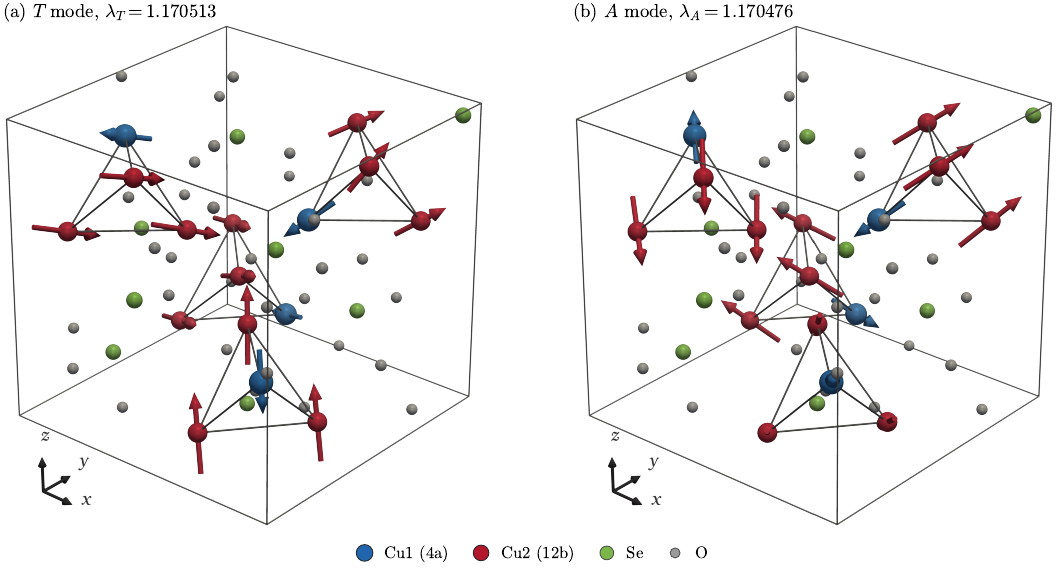}
\caption{\label{fig:cuo}%
Leading $\Gamma$-point response eigenvectors of Cu$_2$OSeO$_3$: (a) the state of the threefold-degenerate $T$ mode whose net moment points along $[111]$ and (b) the nearly degenerate, fully symmetric $A$ mode.
A cubic Cu$_2$OSeO$_3$ cell holds the four Cu$_4$ tetrahedra.
Each tetrahedron has one Cu1 site at the $4a$ Wyckoff position (blue) and three Cu2 sites at the $12b$ position (red).
Arrows indicate the site magnetic moments: in each tetrahedron, for both modes, the blue moment is directed opposite to the sum of the three red moments, corresponding to a $3{:}1$ ferrimagnetic configuration.
In the $A$ mode, the four moments of the tetrahedron are directed along the four $\langle111\rangle$-type directions and mutually cancel out.}
\end{figure*}

\textit{Cu$_2$OSeO$_3$: ferrimagnetic order}---In the chiral cubic insulator Cu$_2$OSeO$_3$ (space group $P2_13$), in addition to the ferrimagnetic state emerging at temperatures below $\approx 60$~K~\cite{BosCu2OSeO3_2008}, a long-wavelength helical structure and a skyrmion lattice can be realized~\cite{SekiCu2OSeO3_2012,AdamsCu2OSeO3_2012}.
Its unit cell, containing 56 atoms, comprises sixteen Cu ions distributed across two inequivalent Wyckoff positions: four Cu1 sites ($4a$) and twelve Cu2 sites ($12b$).
Within each Cu$_4$ tetrahedron, the magnetic moment at the $4a$ site is oriented antiparallel to the moments of the other three ions~\cite{BosCu2OSeO3_2008,JansonCu2OSeO3_2014}.
Since the wavevector of the helical structure is much smaller than those we sample~\cite{suppl}, we analyze the response at the $\Gamma$ point to determine whether it reveals the known magnetic configuration without specifying it beforehand.

For an arbitrary initial state and without imposing symmetry constraints, the calculation of magnetic instabilities at the $\Gamma$ point yields states corresponding to the irreducible representations of the point group $23$~\cite{suppl}.
The dominant mode, with $\lambda_T=1.170513$, corresponds to the triply degenerate representation $T$, the only one possessing a net moment.
The solver returns two of its three partners.
The fully symmetric mode $A$ lies only $3.7\times10^{-5}$ lower, at $\lambda_A=1.170476$~\cite{suppl}.

Integrating site moments over atomic spheres~\cite{suppl} reveals that the moment at the $4a$ position of each tetrahedron is antiparallel to the net moment of the three neighboring $12b$ sites, while the oxygen spheres account for approximately $30\%$ of the total site moments.
In $T$, the tetrahedron moments add up to a net moment, as in the experimentally studied ferrimagnet [Fig.~\ref{fig:cuo}(a)], whereas in $A$ they point along the four $\langle111\rangle$-type axes and mutually cancel out [Fig.~\ref{fig:cuo}(b)].
The obtained response confirms the $3{:}1$ configuration within the strong tetrahedra, a feature characteristic of the first-principles spin model~\cite{JansonCu2OSeO3_2014}, also observed via neutron diffraction in the helimagnetic state~\cite{Franke2019}, but leaves the relative orientation of the tetrahedra open. 
We attribute the near-complete degeneracy of the $T$ and $A$ levels to using the metallic state as a reference.

\begin{figure}[!tbp]
\includegraphics[width=\columnwidth]{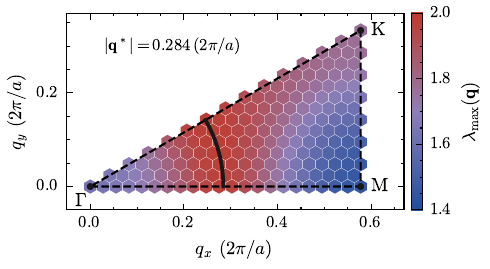}
\caption{\label{fig:nii}%
Leading response eigenvalue $\lambda_{\max}(\mathbf q)\equiv\lambda_{\mathbf q}$ of monolayer NiI$_2$ over the irreducible part of the Brillouin zone, the triangle $\Gamma$--M--K.
Dashed lines mark the $\Gamma$--M--K--$\Gamma$ boundary, and the solid arc marks the fitted mean radius $|\mathbf q^\ast|=0.284\,(2\pi/a)$ of the ridge of large eigenvalues, with $a$ the in-plane lattice constant.
The axes are the in-plane Cartesian components $q_x$ and $q_y$ of $\mathbf q$, and $\Gamma$, M, and K are high-symmetry points.}
\end{figure}

\textit{Monolayer NiI$_2$: helimagnetic order}---Bulk NiI$_2$, consisting of a stack of triangular nickel layers, is a helimagnet in which a spiral magnetic structure gives rise to ferroelectric properties~\cite{Kuindersma1981,KurumajiNiI2_2013} and $p$-wave magnetism~\cite{SongPwave2025}.
The helical, multiferroic order is reported to survive when the bulk is thinned down to a single layer~\cite{AmorosoNiI2_2020,FumegaLadoNiI2_2022,SongNiI2_2022,AminiNiI2_2024}.
Subsequently, spin-polarized scanning tunneling microscopy revealed a spiral structure in the monolayer with a wavevector of $|\mathbf q|=0.220\,(2\pi/a)$, where $a$ is the in-plane lattice constant~\cite{MiaoNiI2_2025}.

Since symmetry does not fix the wavevector, we scan the response for the NiI$_2$ monolayer over the Brillouin zone.
The leading eigenvalues form a ridge around the $\Gamma$ point with a mean radius of $|\mathbf q^\ast|=0.284\,(2\pi/a)$ (Fig.~\ref{fig:nii}) and a maximum of $1.976$ at the point $\mathbf q_{\max}=(\tfrac17,\tfrac17)$ on the $\Gamma$--K line~\cite{suppl}.
Along the ridge, the radius varies by $1.7\%$, while the eigenvalue changes by only $0.54\%$, whereas at the $\Gamma$ point it drops to $1.638$~\cite{suppl}.
This implies that the response constrains the wavevector magnitude much more precisely than its direction.

The obtained wavevector magnitude is consistent with generalized-Bloch spin-spiral calculations excluding SOC~\cite{SodequistOlsen2023}.
However, both approaches overestimate the value by about $30\%$ compared to experiment, whereas supercell calculations incorporating an on-site Coulomb correction reproduce the experimental value accurately~\cite{MiaoNiI2_2025}.
Our maximum lies along the $\Gamma$--K direction, which corresponds to the propagation direction of the helical structure in the monolayer~\cite{MiaoNiI2_2025}.
Competing interactions typically give rise to rings of nearly degenerate wavevectors~\cite{AmorosoNiI2_2020,Bergman2007}.
However, selecting a specific state from the set—or choosing between a spiral and a spin-density wave that have the same wavevector—lies beyond the scope of linear response theory.

\textit{Conclusions}---We have demonstrated that the leading eigenmodes of the self-consistent response of a paramagnetic reference state---calculated from first principles with SOC---determine the magnetic structures of three distinct materials: the stacking type and magnetic moment orientation in KV$_2$Se$_2$O, the $3{:}1$ tetrahedral clusters in Cu$_2$OSeO$_3$, and the helical instability of a NiI$_2$ monolayer.
The instabilities are those of the paramagnetic, here metallic, reference, and their eigenvalues depend on the smearing and on the Brillouin-zone sampling.
Correlated paramagnets with fluctuating local moments need a reference from dynamical mean-field theory~\cite{Kotliar2006}.

The Hartree--exchange--correlation kernel follows from the reference state, free of adjustable interaction parameters, and because it enters only through its action on the four-component density, gradient corrections, a Hubbard correction with $U$ from linear response~\cite{Cococcioni2005,TimrovHubbard2018,TimrovHubbard2021,BinciTimrov2025}, and screened interactions of Bethe--Salpeter type~\cite{OnidaReiningRubio2002} can be added without changing the solver.
Finally, since one eigensolve per wavevector replaces a list of candidate orders, the method can screen databases of nonmagnetic structures for magnetic instabilities beyond catalogs built on measured orders~\cite{GallegoMAGNDATA2016,XuMagneticTopological2020}, and should prove valuable in the search for materials with novel and complex magnetic properties.

\begin{acknowledgments}
We thank Hugo Dil, Iurii Timrov, and Jaroslav Fabian for fruitful discussions.
Y.Z., E.C.-A., and O.V.Y. acknowledge Swiss National Science Foundation Grants No.~224624 and 238150.
Calculations used Swiss National Supercomputing Centre resources under Project No.~lp96 and the facilities of the Scientific IT and Application Support Center of EPFL.
D.K. acknowledges Taiwan National Science and Technology Council Grant No.~114-2112-M-006-034-MY3, Slovak Academy of Sciences Project IMPULZ-2021-26 SUPERSPIN, and Slovak Research and Development Agency Contract No.~APVV-24-0134.

Yaroslav Zhumagulov wrote the response code, its tests, the calculation workflows, and the analysis and plotting scripts with the AI coding assistant Claude Code (Anthropic), using the Claude Opus (versions 4.8, 5, and 5.5) and Claude Fable (versions 5 and 5.1) models, which also searched the literature, checked citations against their sources, and drafted and revised parts of the text.
Yaroslav Zhumagulov directed each step, chose the identities that validate the code~\cite{suppl}, checked the derivations, numerical results, figures, and cited statements, and takes full responsibility for them.
\end{acknowledgments}

\textit{Data availability}---The data supporting our findings were generated by numerical calculations, and we will make the source code, \texttt{magnetic-detector} (MIT license), and the inputs and scripts of all calculations publicly available upon publication.

\nocite{QE2017,PerdewWang1992,HamannONCV2013,PseudoDojo2018,DalCorso2005,Sorensen1992,ARPACK1998,SciPy2020,PerdewBurkeErnzerhof1996,ChenKV2Se2O2025}
\bibliography{refs}

\end{document}